# Social Service Brokerage based on UDDI and Social Requirements

Jan Świerzowicz and Willy Picard

Department of Information Technology, Poznan University of Economics,
Mansfelda 4, 60-854 Poznań
{jswierz,picard}@kti.ue.poznan.pl

**Abstract.** The choice of a suitable service provider is an important issue often overlooked in existing architectures. Current systems focus mostly on the service itself, paying little (if at all) attention to the service provider. In the Service Oriented Architecture (SOA), Universal Description, Discovery and Integration (UDDI) registries have been proposed as a way to publish and find information about available services. These registries have been criticized for not being completely trustworthy. In this paper, an enhancement of existing mechanisms for finding services is proposed. The concept of Social Service Broker addressing both service and social requirements is proposed. While UDDI registries still provide information about available services, methods from Social Network Analysis are proposed as a way to evaluate and rank the services proposed by a UDDI registry in social terms.

**Keywords:** Service Brokerage, UDDI, Social Network Analysis, Social Requirements, Service Oriented Architecture.

## 1 Introduction

The Service Oriented Architecture (SOA) has been defined by OASIS (Organization for the Advancement of Structured Information Standards) as a "paradigm for organizing and utilizing distributed capabilities that may be under the control of different ownership domains" [1]

According to OASIS, the main idea underlying SOA is to help needs to be satisfied by services performed by entities having capabilities. An important issue in SOA is therefore the identification of services fulfilling the requirements of the needy actor. A widely accepted solution to the service search issue is the use of a service broker: in this approach, the needy actor sends its requirements to the service broker who seeks next for a service fulfilling the needed requirements. When such a service has been identified, the service broker informs the needy actor about the service(s) that satisfying the requirements, as presented in Figure 1.

The service broker approach is a common approach in distributed systems, with applications e.g. in the Java world with RMI [2] or in Common Object Request Broker Architecture (CORBA) [3] In SOA, and more precisely in its most common implementation technologies, i.e. Web Services, the service broker approach is based on the concept of UDDI registry in which descriptions of services and their providers are stored.

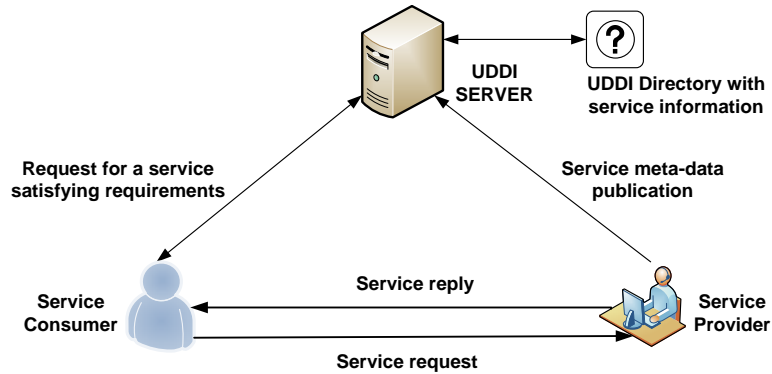

**Fig. 1.** Looking up service using UDDI (adapted from [9])

UDDI, which stands for *Universal Description, Discovery and Integration*, consists of a set of web-services providing access to information about a business or other entity and its technical interfaces (or APIs). UDDI registries are run by Operator Sites. UDDI registries may be used to publish and find information about one or more businesses or entities. As a consequence, UDDI registries allow service consumers to discover available services fulfilling given requirements, with the specification of the technical programming interfaces, usually written in the Web Services Description Language (WSDL) [4]

Lately UDDI has been criticized primarily because of its *decentralized* approach. Service providers are responsible for managing the contents of UDDI registries. UDDI operators do not monitor the entries in their UDDI registries. As a consequence, many entries concern companies that no longer exist, or provide imprecise information about organizations. As a consequence, UDDI registries have not been endorsed massively as foreseen in the first years of 2000's, leading to Microsoft, IBM and SAP discontinuing their UDDI Business Registry (UBR) project for Web services [5].

As a consequence of the decentralized approach of UDDI registries and the strong focus on services, "social" aspects, such as reputation, trust, and previous collaboration history of service providers, are not supported by UDDI registries. However, there is a need, not only for efficient mechanisms for the search of services satisfying a set of functional requirements (as provided by UDDI registries), but also for an evaluation of found services in social terms (answering questions such as "is a service satisfying my functional requirements trustworthy).

The purpose of this paper is to propose a solution for enhanced Service Provider selection taking into consideration social aspects. The concept of *Social Service Broker* is introduced as an intermediary between Service Consumer and Service Provider. The solution proposed in this paper is based on UDDI backed by Social

Network Analysis, which addresses abovementioned social aspects. To our best knowledge, no such solution was proposed earlier.

The paper is organized as follows. In section 2, the features of UDDI registries are briefly introduced. In Section 3, Social Network Analysis and the idea of Social Requirements are presented. In section 4, the concept of Social Service Broker is detailed. Next, a short example illustrating a potential use of a Social Service Broker is given. Finally, Section 6 concludes the paper.

## 2 UDDI

UDDI registries implement the UDDI standard[6], providing transactional access to search and update registry entries. UDDI enables organizations, *service providers*, to publish information about their companies and services. These services and their technical details may then be discovered by other potential trading partners, *service consumers*. Once a service is chosen, the trading partners can exchange request/response messages and thus integrate with each other, potentially realizing extended enterprise solutions in Web services frameworks, allowing heterogeneous platforms and applications to discover and interoperate with existing applications [7]

### 2.1 UDDI Components

The information that a service provider can register encompasses various information "areas", providing answers to the questions "who, what, where and how" about services.

In UDDI, the familiar metaphor of telephone directories is applied to Web services registries: UDDI registries contain White Pages, Yellow Pages and Green entries [7, 8]. UDDI Components are as follows:

**White Pages**
> Contain basic contact data of the service provider (name, address, phone, Web site URL, and description of the organization.

**Yellow Pages**
> A set of categories and taxonomies that classify services and service providers

**Green Pages**
> Contain technical information about services- details about their location and the way they should be invoked

### 2.2 UDDI Registry Data Structures

In the UDDI API Schema [4], a set of UDDI data structures has been specified: business entities, business services, and binding templates.

The *business entity* is the top level UDDI element and represents the organization that is publishing its services. This element contains basic information about organization; i.e. name, business description, contact information, identifiers and classifications for an organization entity and a description of the services the organization is offering.

Each business entity contains a number of *business services*. A business service has a name, description, and categorizations. So by using business services, a business entity can organize its services.

Each service contains a number of *binding templates* that specify where the service is located and how to invoke it. Each binding template contains a set of keys that link the binding template to more detailed technical specifications ("*tModels*") which in turn contain more detailed technical information about the services. tModels can be regarded as "reference standards" and may be re-used by several different business entities [7, 8].

## 3 SNA and Social requirements

### 3.1 Social Network Analysis

A social network is a graph of nodes (sometimes referred as actors), which may be connected by relations (sometimes referred as ties, links, or edges). Social Network Analysis (SNA) is the study of these relations[10].

An important aspect of SNA is the fact that it focused on the how the structure of relationships affects actors, instead of treating actors as the discrete units of analysis. SNA is backed by social sciences and strong mathematical theories like graph theory and matrix algebra [11] , which makes it applicable to analytical approaches and empirical methods. SNA uses various concepts to evaluate different network properties.

Recently, numerous networking tools have been made available to individuals and organizations mainly to help establishing and maintaining virtual communities. The common characteristic to all of them is that members build and maintain their own social networks, which are, then, connected to other networks through hubs (individuals that are members of two or more networks) [12].

### 3.2 Social Requirements

Social Network Analysis may be used to examine a given network by evaluating its properties. Social requirements may be considered as the reverse approach: using social requirements, properties of a network and their associated expected values are defined. One may then check if an existing network satisfies these social requirements. It should be noticed that social requirements are usually at a higher level of abstraction than SNA metrics, and therefore, a "translation" phase between social requirements and SNA metrics is usually required. [13]

### 3.3 Finding Services with Social Requirements

Social Network Analysis may also be used to find a service provider to fulfill the need. Unlike using UDDI registries, provider's selection may be enhanced with the issues of trust and previous collaboration while using SNA and social requirements. Unfortunately, such searching is time consuming because one needs to browse whole social network structure and check if a given actor 1) provides a required service 2) meets social requirements.

## 4 Social Service Broker

The concept of Social Service Broker is proposed in this paper as a combination of the two approaches – UDDI registries and Social Network Analysis. Comparing the search for a service with a Social Service Broker to the search for a service in a human society:
- UDDI registries are similar to phone books that may be consult to find information about companies providing a given service, e.g. plumbing;
- SNA methods may lead to ranking of available services, similarly to the opinion of friends about a given service provider, e.g. "is John Smith a good plumber".

Therefore, combination of UDDI registries and SNA methods is similar to the case when a list of service providers coming from a phone book is then ranked based on the opinion of friends.

In the proposed approach, the Social Service Broker is responsible for both aspects, i.e. retrieving the list of service providers satisfying the requirements of the service consumer, as well as ranking the results from the UDDI registry according to a set of social requirements.

In Figure 2, the processing of a request sent by the Service Consumer to the Social Service Broker is illustrated. In this process, seven steps may be distinguished:
1. Both service and social requirements are specified and sent out by the Service Consumer to the Social Service Broker.
2. Service specifications are sent by the Service Broker to a UDDI registry, which seeks entries in yellow and green pages and generates the list of services satisfying the received requirements
3. Meta-data from the white pages about the providers of services satisfying the received requirements are sent back to the Social Service Broker by the UDDI registry.
4. The Social Service Broker sends a list of service providers from the white pages meta-data sent by the UDDI registry, together with the set of social requirements sent by the Service Consumer to the Social Network Server.
5. The Social Network Server evaluates the list of service providers as regards the social requirements, and sends back the results of its evaluation to the Social Service Broker.
6. The Social Service Broker sends a ranked set of service providers to the Service Consumer.

7. Based on the ranking of found service providers, the Service Consumer makes a choice of the most suitable service provider and sends him a request to perform the required service.

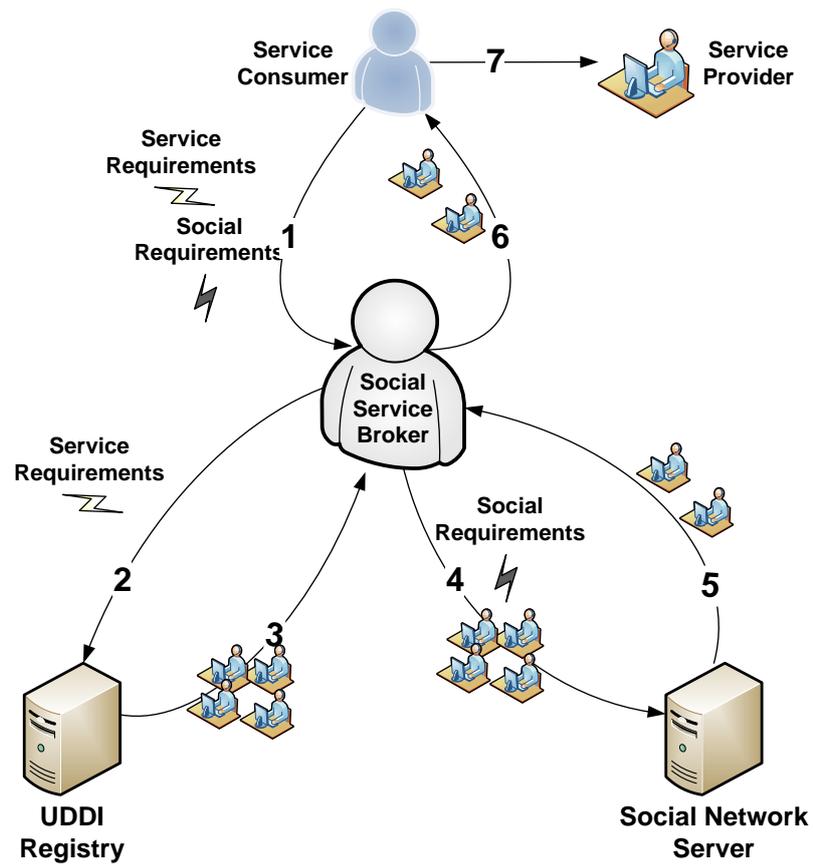

**Fig. 2.** Social Service Brokerage

It should be noted that set of services that satisfies requirements may be empty (if the requirements are too strict); on the other hand it may contain many solutions (if the requirements are too vague).

## 5 Example

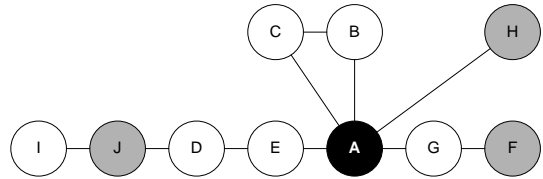

**Fig. 3.** Social Network of Organization A

To illustrate the concepts presented in the former section, let's assume that the partnerships of an organization (A) with other organizations are modeled as the social network presented in Figure 3. Two organizations are or were partners if an edge between these organizations exists in the social network presented in Figure 3.

The organization A has a given **need**, such as the writing of a performance report concerning a new product to highlight new features in an advertisement campaign. Company A sends its social and service requirements to a Social Service Broker. Afterwards the Social Service Broker forwards the service requirements to a UDDI registry and the result is that such report (**a capability to fulfill the need**) can be written by actors F, H, and J.

Along with service requirement, the following social requirement was also specified: a service provider must be a company that either company A or a direct partner of company A has already collaborated with.

As it can be seen from Figure 3, only actors H (direct collaborator) and F (collaboration through actor G) meet the social requirements. However, actor H should be ranked higher than actor F, as actor H was/is a direct collaborator while the collaboration with F took/takes place via the actor G. Therefore, the Social Service Broker should send back a list of two service providers, i.e. companies G and H, satisfying both the service and social requirements, together with the appropriate ranking, i.e. actor H is ranked higher than actor F.

## 6 Conclusions

UDDI registries have been under a lot of criticism lately, partly because UDDI registries do not support social aspects related with the choice of appropriate services and service providers. The main contribution in this paper is the concept of Social Service Broker. To our best knowledge, the integration of Social Network Analysis/Social Requirements with UDDI is a novel approach to service brokerage.

Thanks to the proposed Social Service Broker, actors with **needs** may more easily find actors who have **capabilities to fulfill these needs in a socially satisfying manner,** with support both for efficient search (provided by UDDI registries) and social issues (modeled as social requirements).

Among future works, the concepts presented in this paper should be formally defined. Mechanisms to evaluate the relevance of social properties of the collaborative partners are still to be proposed. Additionally, within the context of the IT-SOA project [14], a prototype implementing the concept of Social Service Broker is currently under development. The project will gather companies from the construction sector in the Great Poland region. During the IT-SOA project, the Social Service Broker is planned to be a core element of the Service-Oriented infrastructure that construction companies will use to integrate their processes, leading to an evaluation of the proposed Social Service Broker concept in a real business environment.